\documentclass[journal]{IEEEtran}

\usepackage{graphicx}
\usepackage{color}
\usepackage{placeins}
\usepackage{float}
\usepackage{tabularx,colortbl}
\usepackage{amssymb}
\usepackage{amsthm}
\usepackage{cite}
\usepackage{amsmath}
\usepackage{makecell}

\usepackage{caption2}

\usepackage{algorithm}
\usepackage{algpseudocode}

\theoremstyle{plain}

\theoremstyle{plain}

\IEEEoverridecommandlockouts

\begin{document}

\title{Mobile Cell-Free Massive MIMO: Challenges, Solutions, and Future Directions}

\author{Jiakang Zheng, Jiayi Zhang, Hongyang Du, Dusit Niyato,~\IEEEmembership{Fellow,~IEEE}, Bo Ai,~\IEEEmembership{Fellow,~IEEE}, M{\'e}rouane~Debbah,~\IEEEmembership{Fellow,~IEEE}, and Khaled B. Letaief,~\IEEEmembership{Fellow,~IEEE}

\thanks{J. Zheng, J. Zhang, and B. Ai are with the School of Electronic and Information Engineering and Frontiers Science Center for Smart High-speed Railway System, Beijing Jiaotong University; H. Du and D. Niyato are with Nanyang Technological University; M. Debbah is with Khalifa University of Science and Technology; K. B. Letaief is with Hong Kong University of Science and Technology.}
}

\maketitle
\vspace{-1.8cm}
\begin{abstract}
Cell-free (CF) massive multiple-input multiple-output (MIMO) systems, which exploit many geographically distributed access points to coherently serve user equipments via spatial multiplexing on the same time-frequency resource, has become a vital component of the next-generation mobile communication networks. Theoretically, CF massive MIMO systems have many advantages, such as large capacity, great coverage, and high reliability, but several practical obstacles must be overcome. In this article, we study the paradigm of CF massive MIMO-aided mobile communications, including the main deployment architectures and associated application scenarios. Furthermore, we thoroughly investigate the challenges of mobile CF massive MIMO communications. We then exploit a novel predictor antenna, hierarchical cancellation, rate-splitting and dynamic clustering system for mobile CF massive MIMO. Finally, several important research directions regarding mobile CF massive MIMO communications are presented to facilitate further investigation.
\end{abstract}


\IEEEpeerreviewmaketitle

\section{Introduction}

The continuous upgrading of wireless mobile communications and enhancement of key performance indicators such as higher rate, more robust reliability, and lower latency have never stopped. However, the achievable performance would be negatively affected by the mobility of terminal devices due to the Doppler frequency shift, time-varying channels, frequent handovers, etc. Recently, sixth generation (6G) has started to be researched and is expected to give extended 5G capabilities to support the massive Internet of Things devices and enable seamless operation with a wide range of service requirements \cite{8808168}. In 6G communications, various mobility scenarios and mobility devices will be a part of future wireless communication networks, e.g., vehicle communication, high-speed train (HST) communication, unmanned aerial vehicle (UAV) communication, robotic communication, satellite communication, etc. To meet the increasing demand for mobility, it is necessary to be aware of the advancements in innovative architectures, breakthrough technologies, and their challenges.

Recently, massive multiple-input multiple-output (MIMO) systems have attracted increased research attention and shown their powerful potential for high-mobility communications, particularly with rich scatterers because of their superior spatial resolution. By exploiting a large number of antenna elements, the base stations (BSs) of centralized massive MIMO systems may use beamforming to create high-gain and narrow beam targets that can accurately track fast-moving terminals. Moreover, narrow beams can also be used to achieve multiuser spatial division multiple access, hence increasing the total capacity of the mobile communication system. However, due to the high mobility, fast signal processing limited by matrix dimensionality and frequency handover between neighboring BSs are two of the primary challenges for the practical implementation of massive MIMO into mobile communications system design. In addition, the poor energy efficiency (EE) of the cellular massive MIMO system is also a critical concern.

One promising solution for addressing these core difficulties is to use a distributed massive MIMO architecture known as cell-free (CF) massive MIMO \cite{9586055}. Through spatial multiplexing on the same time-frequency resource, each user equipment (UE) in the CF massive MIMO system is coherently served by a large number of geographically distributed access points (APs) that are connected to a central processing unit (CPU) \cite{kanno2022survey}. As a result of the user-centric paradigm, the UE is blind to the cell boundary and can thus obtain seamless and uniform coverage when moving \cite{9650567}. Obviously, the architecture of CF massive MIMO systems can reduce handover delay and handover failure probability, which satisfies the requirement of high mobility.
Besides, the distributed architecture of CF networks enhances the likelihood of having line-of-sight paths and improves the angular resolution, thus facilitating more precise positioning of mobile UEs.
In particular, the architecture of CF massive MIMO is suited for distributed processing, which can replace complex centralized processing and achieve accelerated computing of large amounts of data.
Moreover, the CF massive MIMO system features both large-scale antenna gain and macro-diversity gain, as well as low path loss, enabling it to improve the spectral efficiency (SE) and EE of mobile communications by an order of magnitude in comparison to the cellular massive MIMO \cite[Fig.~8]{8097026}.

\begin{table*}[th]
\centering
\caption{An overview of mobile CF massive MIMO communication systems \cite{8808168,8097026,9586055,kanno2022survey,9650567}.}
\vspace{3mm}
\includegraphics[scale=0.52]{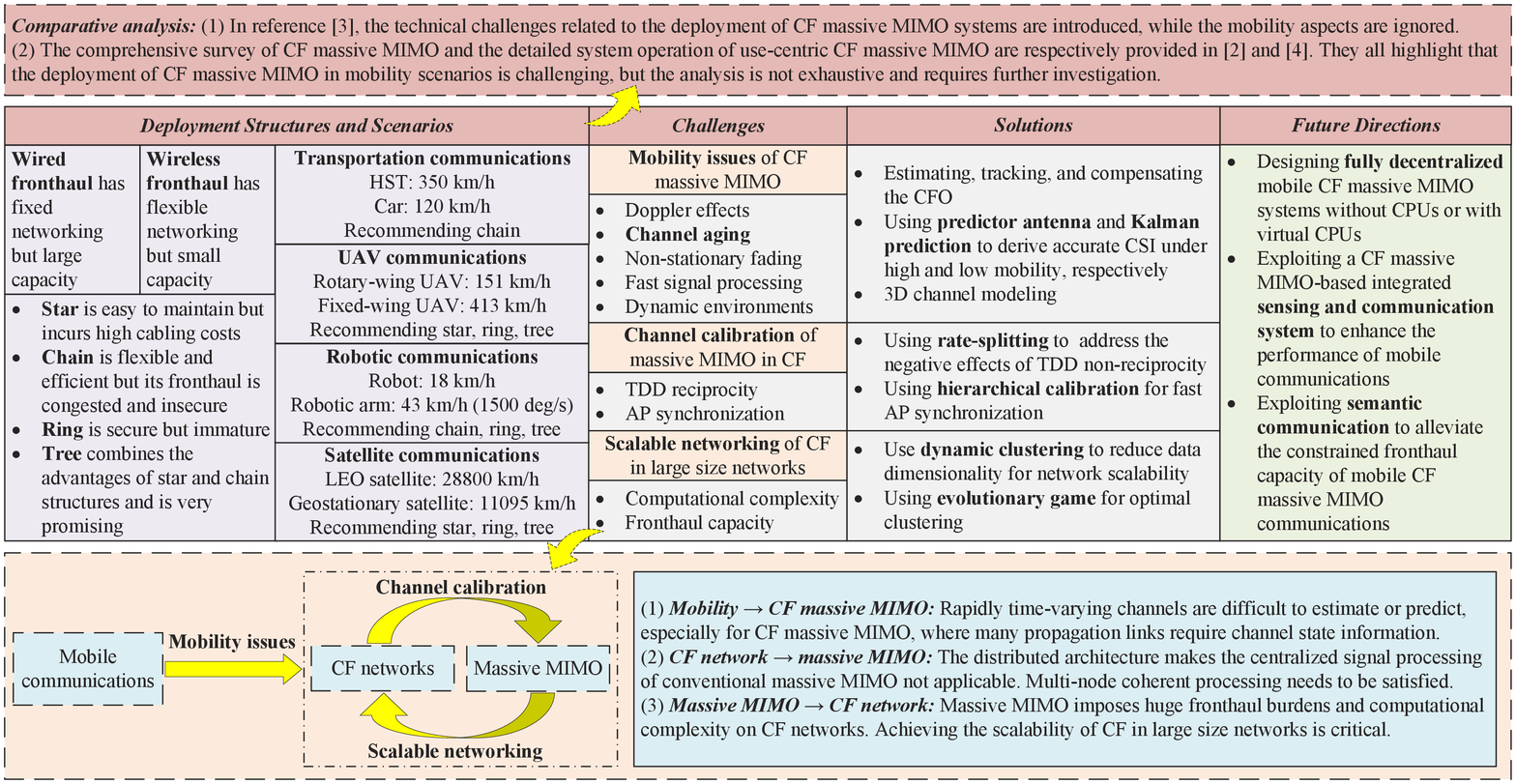}
\vspace{-3mm}
\label{fig:table}
\end{table*}

Although CF massive MIMO-aided mobile communication systems offer several potential advantages, they are also facing many challenges in actual deployment. Specifically, the following questions need to be answered:
\begin{itemize}
  \item [\textbf{Q1)}]
  How CF massive MIMO is deployed in mobile communications and serves mobile UEs?
  \item [\textbf{Q2)}]
  What are the main challenges in the practical application of mobile CF massive MIMO systems?
  \item [\textbf{Q3)}]
  Which technologies can solve the challenges in \textbf{Q2} and how do they differ between CF and cellular network?
\end{itemize}

This article will first provide an overview of the mobile CF massive MIMO communication system as shown in TABLE~\ref{fig:table}.
The main contributions are summarized as follows:
\begin{itemize}
\item We provide a comprehensive discussion about the applications of mobile CF massive MIMO communications, serving transportation, UAV, robotics, and satellite. Then, we propose four deployment structures based on features and requirements of various application scenarios.
\item We investigate the challenges of the practical application of mobile CF massive MIMO communication systems, including mobility issues, channel calibration, and scalable networking. Besides, we present promising solutions, such as predictor antenna, rate-splitting, and dynamic clustering, and compare their differences under mobile CF and cellular networks.
\item Importantly, we propose a fast and low-complexity hierarchical calibration framework for mobile CF massive MIMO communication systems. Besides, we also propose an efficient evolutionary game-based clustering scheme. Then, numerical results validate the performance of the proposed solutions. Finally, a few novel research directions are highlighted.
\end{itemize}

\section{Deployment Structures and Scenarios}

\begin{figure*}[t]
\centering
\includegraphics[scale=0.62]{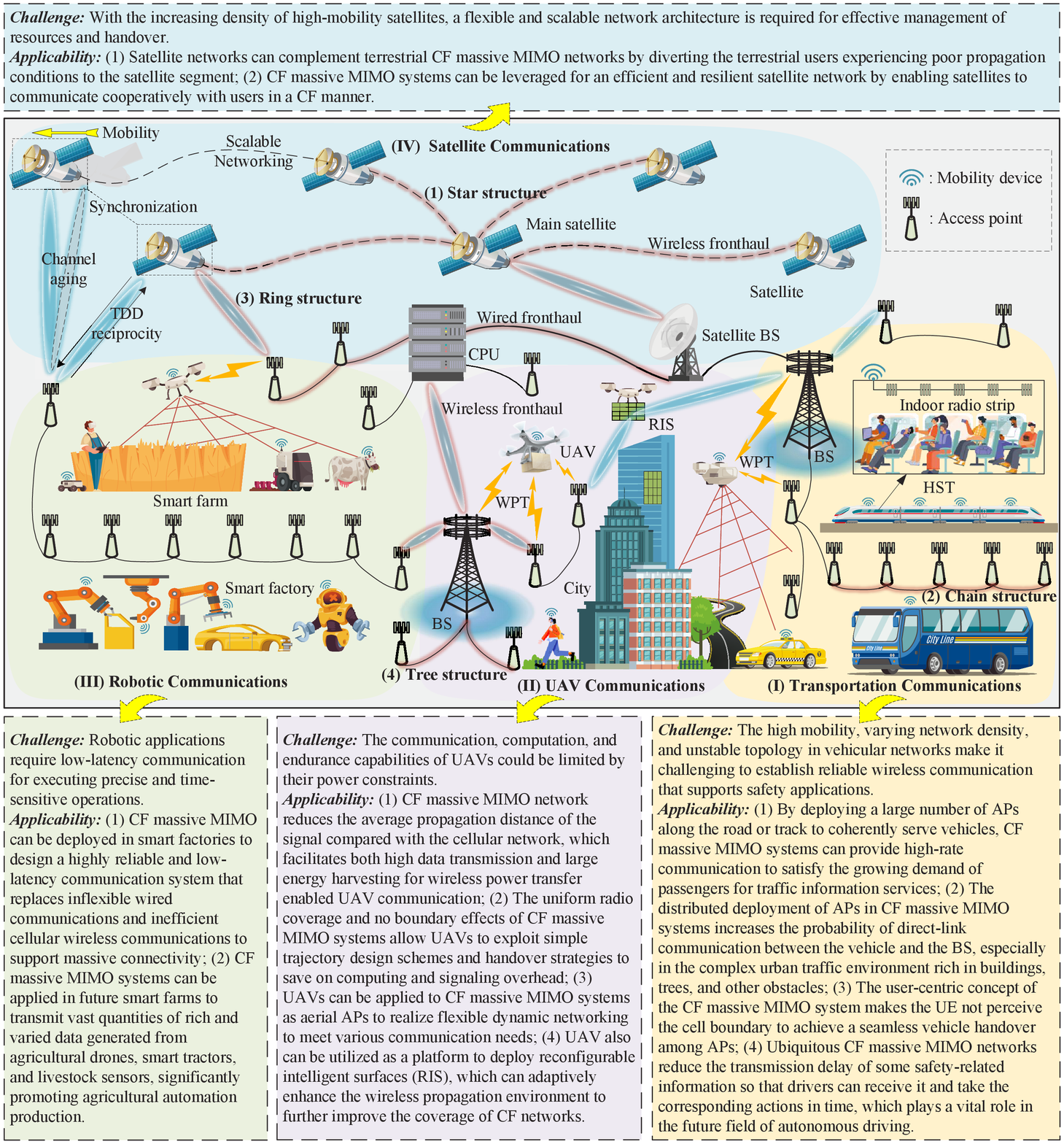}
\caption{Deployment structures and scenarios of mobile CF massive MIMO communication systems.}
\vspace{-3mm}
\label{fig:CFmobile}
\end{figure*}

We first propose four typical deployment structures for mobile CF massive MIMO communications, as shown in Fig.~\ref{fig:CFmobile}, namely star, chain, ring and tree structures.
The specific strengths and weaknesses of each option are outlined below.
\begin{itemize}
\item \textit{Star:} Most of the current research on CF massive MIMO systems uses the star structure. Specifically, each AP is directly wired to the CPU, which has the advantage of easy maintenance and equipment replacement. However, the number of AP connections in the star structure is limited by the number of CPU interfaces. Moreover, it can result in significant costs related to cabling.
\item \textit{Chain:} In order to overcome the shortcomings of the star structure, the chain structure is proposed by letting each AP connect to the CPU in a cascading manner. This enables more flexible and efficient deployment of CF massive MIMO systems, especially in indoor environments. Obviously, the chain structure will cause high fronthaul requirements, which also limits the number of APs on the chain. In fact, the security of the chain structure is poor because most APs will stop serving when the connection close to the CPU fails.
\item \textit{Ring:} To address issues in the chain structure, a ring structure can be used to provide an alternate transmission path, which achieves highly reliable networks at lower fronthaul requirements. However, theoretical research on the ring structure is still immature, and its application in practical scenarios still needs further approach.
\item \textit{Tree:} The tree structure is a combination of the star and chain structures, while also inheriting their advantages. It is very suitable for upgrading existing cellular networks, where the original BS can act as a super AP in CF massive MIMO networks.
\end{itemize}
Note that their connections can be wired or wireless fronthaul.
Compared with the large-capacity wired fronthaul with a fixed topology, wireless fronthaul provides a flexible topology but suffers from its limited capacity. Moreover, the capacity of wireless fronthaul in mobile CF networks is variable due to the changes in APs' positions and interference levels.
Fortunately, the emergence of large antenna arrays that rely on high-frequency bands has made it possible for mobile CF networks to achieve high-capacity wireless fronthaul.

Next, four types of mobile application scenarios for CF massive MIMO systems are introduced, namely transportation communications, UAV communications, robotic communications, and satellite communications. As shown in Fig.~\ref{fig:CFmobile}, we outline the specific challenges related to each scenario and present the applicability of CF massive MIMO in mobility scenarios that conventional wireless networks lack.
Based on the unique features of each scenario, appropriate CF network deployment structures can be selected.
For instance, when UAVs and satellites are used as APs for networking, it is suitable to adopt star, ring and tree structures. For places with regular shapes such as factories, farms, roads and railway tracks, deploying with a chain structure is very efficient. Moreover, for some hot spots, the tree structure is suitable for high-density deployment.

\section{Challenges of CF massive MIMO Implementation}

In this section, we discuss the three key challenges of the implementation of the mobile CF massive MIMO communication system, as shown in TABLE~\ref{fig:table}.

\subsection{Mobility Issues}

Mobility has long been an essential study topic in developing wireless communication systems. One of the primary characteristics of mobile communications is the rapid time-variation of the fading channel induced by the significant Doppler spread. Especially in high-speed railway communication, e.g., with a train speed of $300$ km/h at a carrier frequency of $1.8$ GHz, whose maximum Doppler frequency reaches $500$ Hz and the approximate channel coherence time is only $375$ $\mu$s. Additionally, variations in the speed of mobility devices in real-time will result in time-varying Doppler spreads and non-stationary fading coefficients, which makes accurate modeling and analysis of CF massive MIMO systems in mobility scenarios more difficult. Moreover, CF massive MIMO systems offer multiple-connection communications for each mobility device, making them more susceptible to Doppler effects than conventional single-connection cellular networks.

In addition, the fast time-variation in the channel brought on by device mobility makes it exceedingly difficult to obtain accurate channel state information (CSI), which is also called channel aging. Specifically, due to the estimated channel information becoming outdated after a short time slot, the increased channel estimation error makes it challenging to design high-quality transceivers for subsequent data transmissions. In practice, efficient transceiver design to eliminate inter-user interference contributes significantly to the improvement of service quality in CF massive MIMO systems. Therefore, the performance of CF massive MIMO systems will degrade over time in time-varying channels.

Additionally, mobility-induced Doppler frequency shift results in carrier frequency offset (CFO), which further introduces inter-carrier interference and seriously impairs the performance of multi-carrier CF massive MIMO systems.
One approach is to estimate, track and compensate CFO, which can be challenging since its time-varying nature due to the Doppler shifts changing over time. Therefore, the temporal correlation of CFO changes needs to be exploited to enhance system operation.
In addition, high mobile communication necessitates that the CF massive MIMO system be able to do fast signal processing by designing low-complexity and high-efficiency algorithms, such as pilot assignment, power control, transceiver design, etc. Moreover, dynamic changes in scattering clusters due to the random movement of UEs, and near-field propagation characteristics brought about by the short transceiver distance of CF networks make the conventional channel model no longer applicable to CF massive MIMO mobile communication systems. Besides, different AP deployment environments (e.g., altitude, direction and blockage) will produce different channel characteristics, so that the channel matrix of the entire network has different statistics.
Therefore, accurate and general three dimensional (3D) channel modeling is required to ensure the accuracy of obtained conclusions.

\subsection{Channel Calibration}\label{se:ca}

CF massive MIMO systems generally use the time-division duplex (TDD) protocol to learn the downlink channel from the uplink pilot signal, which relies on channel reciprocity. Additionally, APs need to maintain a relatively accurate time and phase synchronization to provide coherent transmission in a CF massive MIMO system consisting of many geographically distributed APs. Therefore, TDD reciprocity calibration and AP synchronization calibration are two key issues for the practical application of CF massive MIMO systems \cite{6760595}.

Notably, the assumption of channel reciprocity allows the AP to accurately determine the precoding weights needed for downlink data transmission using the known uplink channel information. Although the physical propagation channel is purely reciprocal, the effective end-to-end channel also includes the transceiver, whose reciprocity is destroyed by the frequency response of the hardware. Fortunately, the change in the frequency response of the transceiver hardware mainly depends on the operating conditions, e.g., temperature drifts. Therefore, the TDD non-reciprocity characteristics vary slowly compared with the changes in the physical propagation channel. As a result, it is possible to estimate and calibrate the system so that channel reciprocity is maintained with negligible overhead.
There are numerous established calibration methods at present, such as measuring the uplink and downlink channels twice, inverse calibration, calibrating with machine learning, etc. \cite{8411461}.
However, applying these methods to CF massive MIMO systems needs too much processing and feedback overhead because the amount of CSI increases as the number of APs grows. Besides, it is also challenging to calculate the calibration matrix for each AP simultaneously. Thankfully, if the calibration errors are independent across APs, they are averaged in the large system limit \cite{7892949}.

Furthermore, the synchronization between the APs, which is necessary for the joint coherent processing of the APs, considerably guarantees the performance gain in the practical application of the CF massive MIMO system. However, geographically distributed APs induce signal arrival delay differences, which cause delay phases, making coherent transmission challenging. Especially in mobility scenarios, the delay phase of multiple propagation links is more difficult to evaluate and track. Moreover, asynchronous issues are made worse by imperfect transceiver hardware, which can add random phase shifts to the channel.
Although the classic synchronization calibration methods, including the total least-square (TLS) calibration algorithm and the Argos calibration algorithm \cite{cao2022experimental}, can be directly applied to CF massive MIMO systems, they will have a huge computational overhead. The reason is that when the number of APs increases, the dimensionality of the calibration matrix increases as well.

\begin{table*}[th]
\centering
\caption{Comparison of key technologies applied to CF and cellular networks \cite{9760033,cao2022experimental,7470942,9720158}.}
\vspace{3mm}
\includegraphics[scale=0.63]{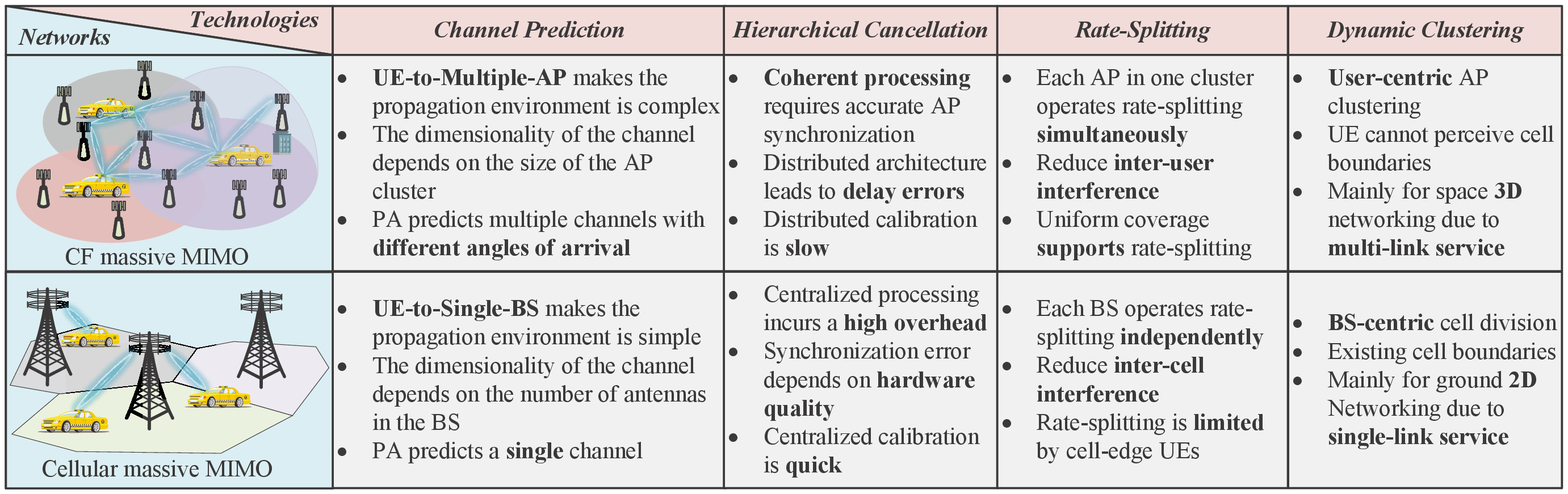}
\label{fig:comparison}
\vspace{-3mm}
\end{table*}

\subsection{Scalable Networking}

The CF massive MIMO system was initially designed to solve the inter-cell interference issue in conventional cellular networks by requiring all APs to serve all UEs simultaneously. However, this is not practical as the size of the network grows since the signal power of distant APs is substantially weakened after reaching the UE, making its service exceedingly inefficient.
Besides, for each AP, the computational complexity of network functions and the fronthaul capacity required for data transmission increase linearly as the number of UEs increases. As a result, the CF massive MIMO system in its initial version was not scalable \cite{9064545}.
A potential approach is to adopt a user-centric deployment by AP clustering, where each UE selects a set of APs that provide the best service conditions.
Each AP must collaborate with different APs while servicing different UEs because these AP subsets differ for each UE.
Additionally, mobility scenarios require UEs to update their serving cluster frequently, making it more challenging for APs to synchronize their network states and perform resource allocation jointly.
At the moment, dynamic clustering is a well-established user-centric cooperation paradigm where the AP subsets can be modified to time-variant features, such as UE locations, service requirements, interference conditions, etc. Furthermore, many processing algorithms can achieve scalability based on the architecture of dynamic clustering, including scalable power control, scalable joint initial access, scalable pilot assignment, scalable uplink combining, and scalable downlink precoding. However, this saving in computational complexity and fronthaul capacity comes at the expense of system performance because each UE is served by fewer APs. Therefore, it is essential and exceedingly challenging to design an advanced AP selection strategy to choose the optimal AP subset for each UE with negligible performance loss.

\section{Promising Solutions}

In this section, we present potential solutions in response to the above challenges.
A comparison of key technologies applied to CF and cellular networks is shown in TABLE~\ref{fig:comparison}.

\subsection{Channel Prediction}

\begin{figure}[h]
\centering
\includegraphics[scale=0.55]{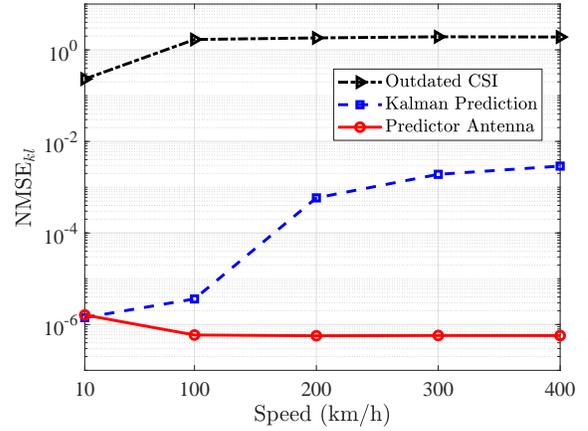}
\caption{NMSE against varying speeds with different prediction methods at SNR = 30 dB.
The carrier frequency, sampling duration, and angle of arrival are $f_c$ = 2 GHz, $T_s$ = 67 $\mu$s, and $\theta_{kl} = \pi/2$, respectively.
For exploiting the memoryfull property of the channel, an AR model of order $M_\text{AR}=10$ in state space form is employed to generate correlated Rayleigh fading processes, as explained in \cite[Appendix C]{9760033}. We benchmark the outdated CSI scheme, where the channel information estimated by the minimum mean square error method gets outdated over time.
\label{fig:PA}}
\end{figure}

Various CSI prediction methods have recently been designed to address the channel aging issue, where mobility causes the CSI to become inaccurate quickly, further degrading the subsequent signal processing capability.
An effective solution is to use the memory property of the channel to predict CSI in advance.
For example, authors in \cite{9210016} developed the channel predictors based on the Kalman filter and machine learning, respectively. Relevant conclusions show that the predictor based on machine learning can approach the predictor based on the Kalman filter with low complexity by conducting offline training. Sadly, their effectiveness is limited to the low speed range and small prediction horizon. For instance, Kalman prediction supports around $0.1-0.3$ wavelengths at $2.68$ GHz and a velocity of $45-50$ km/h \cite{9502657}. A novel prediction paradigm known as the predictor antenna is presented to address the demand for a broader prediction horizon that grows with both velocity and carrier frequency. In such a system, a pair of linearly arranged antennas are installed on the roof of the mobility device, with the predictor antenna mounted in front predicting the channel for the main antenna positioned behind the predictor antenna once the main antenna arrives at the same location several time periods later.

Fig.~\ref{fig:PA} illustrates the normalized mean square error (NMSE) against varying speeds with different prediction methods at SNR = 30 dB. It is clear that when the speed exceeds $100$ km/h, the NMSE of the outdated CSI scheme is larger than 1, which is intolerable in the system's design. Besides, the NMSE of the Kalman prediction scheme increases with the increase in speed, but it significantly outperforms the outdated CSI scheme in every velocity bin. On the contrary, the NMSE of the predictor antenna scheme decreases with the increase in speed, and it starts to be better than the Kalman prediction scheme when the speed exceeds $10$ km/h. Furthermore, their NMSEs become four orders of magnitude different at a speed of $400$ km/h. It is worth mentioning that the high accuracy of the predictor antenna scheme depends on the strong correlation between the predictor antenna and the main antenna, while poor antenna correlation mainly occurs at low velocities \cite{9760033}.

\subsection{Hierarchical Cancellation and Rate-splitting}

\begin{figure}[h]
\centering
\includegraphics[scale=0.55]{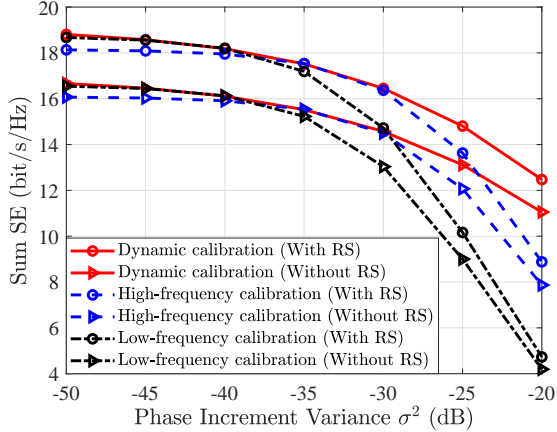}
\caption{Sum downlink SE for CF massive MIMO systems with different phase increment variances. We consider the three-slope propagation model in the simulation setup where $L = 40$ APs and $K =8$ UEs are uniformly and independently distributed within a square of size $100$ m $\times$ $100$ m. Each UE and AP has a single antenna and $N =2$ antennas, respectively. The carrier frequency, bandwidth, and data transmission power are $f_c$ = 2 GHz, $B$ = 20 MHz, and $p$ = 200 mW \cite{8097026}. The oscillator phase of each AP and UE changes with channel use and is modeled by the discrete-time Wiener process.
\label{fig:SE_Phase}}
\end{figure}

As discussed in Section \ref{se:ca}, directly implementing classical calibration methods in CF massive systems will result in significant feedback and computational overhead. A feasible solution to address this issue is to execute hierarchical calibration on the CF massive MIMO system.
The specific steps are as follows:
\begin{itemize}
\item{\textit{Step 1:}} Each AP transmits the measured CSI of each UE to the CPU. Then, the CPU determines the serving AP set for each UE based on the received CSI, thus forming different AP clusters. Actually, APs can also be clustered according to the type of scenario.
\item{\textit{Step 2:}} Each cluster selects an anchor AP with the best CSI. The calibration pilots are then assigned to each anchor AP, with the option of using the same pilot for anchor APs in clusters without intersections \cite{6760595}. Besides, classical calibration methods can be used for calibration between anchor APs to achieve inter-cluster calibration.
\item{\textit{Step 3:}} Dividing the APs in the cluster into two groups, and the two groups of APs should cross as much as feasible. The intra-cluster calibration is then finished using the averaged Argos calibration algorithm \cite{cao2022experimental}.
\end{itemize}
After finishing the three steps above, we can realize the calibration of the entire network.
Clustering and grouping minimize the dimensionality of signal processing and the amount of data exchanged, which lowers computational and feedback overhead and allows for fast and effective calibration of CF massive MIMO systems.

In addition, new signal processing paradigms can also help with the asynchronous problem. As an illustration, channel non-reciprocity causes CSI inaccuracy to lead to a multi-user interference issue, which is the main bottleneck of CF massive MIMO systems. Interestingly, the rate-splitting approach has recently received a lot of attention for its ability to efficiently overcome the interference problem caused by imperfect CSI and also be able to address the negative effects of non-synchronization on CF massive MIMO systems. The rate-splitting strategy for multi-user downlink is based on splitting the message for each UE into a private message and a common message, then combining all the common sub-messages into a super common message, and finally transmitting all the messages simultaneously using superposition coding \cite{7470942}. Each UE at the receivers decodes the common stream first, considering all private streams as noise. The decoded common stream is then removed from the received signal by successive interference cancellation. After that, each UE decodes its own private stream by treating other UEs' private streams as noise. It is emphasized that the uniform coverage properties of CF massive MIMO systems enable rate-splitting to gain considerable performance improvements with a simple and low-complexity common precoding. Importantly, it is shown in \cite{7470942} that rate-splitting makes it possible to significantly reduce the feedback overhead.

We then define three different frequency phase calibration schemes. High-frequency calibration refers to a calibration every 80 channel uses, low-frequency calibration refers to a calibration every 200 channel uses, and dynamic calibration means that its phase calibration frequency is determined by the degree of phase error. Then, as shown in Fig.~\ref{fig:SE_Phase}, we plot the sum SE for CF massive MIMO systems with different phase increment variances. It is evident that the sum SE performance of the CF massive MIMO system falls off significantly as the oscillator phase variance grows. Especially in the case of low-frequency calibration, the sum SE performance loss is the largest, such as it can be reduced by 75\% when the oscillator phase variance varies from $-50$ dB to $-20$ dB. Besides, due to higher time-frequency resources being used by pilot calibration in the case of high-frequency calibration, which performs poorly under low phase increment variance. Furthermore, the dynamic calibration scheme outperforms both high-frequency calibration and low-frequency calibration in every phase increment variance bin due to its ability to adjust the calibration frequency in real time according to the phase increment variance. It is also found that the use of rate-splitting can considerably boost the sum SE performance since a portion of the interference caused by phase error is broadcast so that all UEs may decode and cancel it.

\subsection{Dynamic Clustering}

\begin{figure}[h]
\centering
\includegraphics[scale=0.55]{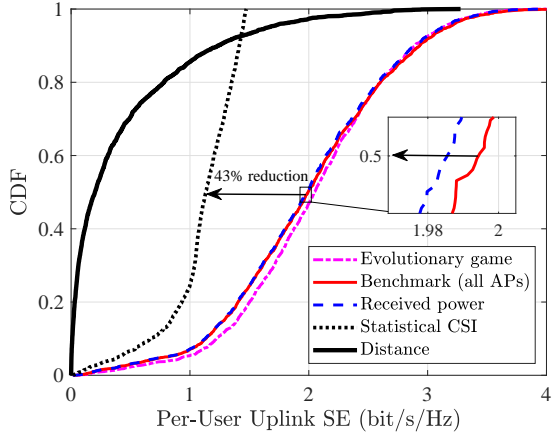}
\caption{CDF of per-user uplink SE of CF systems under different clustering methods. The simulation parameters are the same as in Fig.~\ref{fig:SE_Phase}, except that the number of UEs is $K = 4$ and the setup area size is $500$ m $\times$ $500$ m.}
\label{fig:Clustering}
\end{figure}

In the previous literature, various methods are utilized to research on dynamic clustering with the purpose to achieve scalable CF massive MIMO, such as based propagation distance and statistical CSI.
However, they ignore information such as mobility issues, fronthaul capacity constraints, shadow fading, spatial correlation and estimation error, which will cause non-negligible performance loss.
Game theory is one practical approach to model non-cooperative and cooperative decision making processes in dynamic clustering problems~\cite{9720158}. Different from other game theories, players in evolutionary games are deemed irrational, which is consistent with the reality of dynamic clustering. Specifically, UEs in various locations would select from available APs and modify their selection strategies based on the information obtained online. As a criterion for evaluation, the uplink SE is utilized. Utilizing replicator dynamics to model the evolution of strategy adaptation and achieve evolutionary equilibrium. Consequently, the evolutionary game of dynamic clustering in the CF massive MIMO system can be described as follows.
\begin{itemize}
\item {{\textit{Player:}}} A player is the UE that has more than one choice of AP selection in the service areas. We consider UEs with multiple accessible APs as players. The play can select APs to form a cluster.
\item {{\textit{Population:}}} The population is formed by the set of players in CF massive MIMO service areas. Populations are assumed to be finite.
\item {\textit{Strategy:}} The strategy of each player is the decision of dynamic clustering. In a service area that has $K$ UEs, the set of strategies can be denoted by ${\mathcal{A}_1,\ldots,{\mathcal{A}_K}}$, where $\mathcal{A}_i$ is the clustering scheme of the $i_{\rm th}$ UE.
\item {{\textit{Payoff:}}} The payoff of each player is the utility with given clustering scheme, which is determined by network characteristics. Here, we consider the utility is the SE.
\end{itemize}
We then propose a clustering method based on the formulated evolutionary game. The specific steps are as follows:
\begin{itemize}
\item{\textit{Step 1:}} Initialize the dynamic clustering strategy by establishing the proportions of UEs deciding each clustering scheme.
\item{\textit{Step 2:}} Calculate and compare the players' payoffs (utilities) with the average utility in the service area, considering the utility as the SE.
\item{\textit{Step 3:}} According to the replicator dynamics theory, players will alter their clustering scheme if their SE is below the average SE.
\item{\textit{Step 4:}} Repeat the procedure until the game reaches an equilibrium of evolution.
\item{\textit{Step 5:}} The strategy at the equilibrium point of evolution is the solution to the dynamic clustering issue.
\end{itemize}

As shown in Fig.~\ref{fig:Clustering}, we compare the CDF of per-user uplink SE of CF systems under different clustering methods. For a fair comparison, we assume that each user-centric cluster contains a fixed number of APs. Besides, the case without clustering is provided as a benchmark. It is clear that the propagation distance-based clustering method has the worst SE performance because of its disregard for shadow fading and spatial correlation. Although the statistical CSI-based clustering method can make up for the above defects, it still has a median SE loss of 43\% due to estimation errors.
Moreover, the received power-based clustering method can achieve near-benchmark SE performance by selecting APs based on maximizing the received desired signal power \cite{8097026}.
Interestingly, the evolutionary game-based clustering method can effectively remove negative APs for each UE, making the performance exceed the benchmark without clustering.

\section{Future Directions}

\subsection{Fully Decentralized Network}

Given the variety of CF structures in mobility scenarios and the growing range of services demanded by mobile terminals, multi-CPU control instead of a single CPU becomes the key to realize scalable deployment. However, there will still be a CPU-centric virtual cell boundary concept, and UEs will face the problem of CPU selection/switching during fast movement. Therefore, the connection between CPUs becomes necessary, just like cooperation between APs. The fusion of CPU and AP functions makes it inevitable to design a fully decentralized (without/virtual CPU) mobile CF massive MIMO system. This change also makes the CSI sharing more flexible, such as unidirectional transmission and converged transmission, making the network more resilient to attacks. Moreover, resource allocation and signal processing under fully decentralized mobile CF networks are also important open research issues.

\subsection{Integrated Sensing and Communications}

The distributed architecture of CF massive MIMO shortens the distance between the AP and UE and reduces the path loss, so it naturally supports millimeter wave and terahertz for high-capacity communication. In turn, the application of high-frequency bands also enhances the effectiveness of the large antenna arrays employed by APs. The combination of these benefits creates the potential for enabling mobile CF-based high-gain narrow beams for accurate target tracking and high-capacity wireless fronthaul. Given the overlaps in the spectrums of wireless communication and radar sensing, designing a CF-based integrated sensing and communication system to enhance the performance of mobile communications is also an important research direction. Importantly, the CF network's capability for high line-of-sight transmission and coherent processing can provide further improvements to the sensing capacity for fast-moving targets.

\subsection{Semantic Communications}

The APs in CF massive MIMO systems can be made more functional by implementing semantic communication. Instead of transmitting the original source data, semantic communication uses the semantic information of the data to compress or encode it, thereby reducing the amount of data that must be transmitted to reduce the necessary bandwidth and power consumption. Moreover, encoding the data semantically facilitates the receiver's ability to interpret and utilize the information, thereby increasing the system's dependability. The semantic information of the data to be transmitted can be considered in the physical layer design solutions of CF massive MIMO systems. An exciting research direction is the design of dynamic schemes in CF massive MIMO systems depending on the importance of semantic information. Specifically, the semantic extraction can be done in the CF system, then the data that has more critical semantic information can be prioritized in the wireless transmission. With the cross-layer design of the physical and network layers, the quality of service can be maximized for users despite resource constraints in the CF massive MIMO system.

\section{Conclusions}

In this article, the great potential of CF massive MIMO as a major enabling technology for the next-generation mobile communication system is discussed. We provided an overview of the mobile CF massive MIMO communication system, focusing on four deployment structures and four application scenarios. The main challenges in practical deployment are then discussed, namely mobility issues, channel calibration, and scalable networking. Moreover, respective solutions are provided, namely channel prediction, hierarchical cancellation and rate-splitting, and dynamic clustering. Finally, we presented promising future research directions, which are fully decentralized network, integrated sensing and communications, and semantic communications. We hope that this article can provide useful guidance in designing and implementing CF massive MIMO for next-generation mobile communications.

\vspace{0cm}
\bibliographystyle{IEEEtran}
\bibliography{IEEEabrv,Ref}

\section*{Biographies}

\textbf{Jiakang Zheng} is currently pursuing the Ph.D. degree at the School of Electronic and Information Engineering, Beijing Jiaotong University, Beijing, China. His research interests include cell-free massive MIMO and performance analysis of wireless systems (e-mail: jiakangzheng@bjtu.edu.cn).

\textbf{Jiayi Zhang} [SM'20] is a Professor with the School of Electronic and Information Engineering, Beijing Jiaotong University, Beijing, China. His research interests include cell-free massive MIMO and reconfigurable intelligent surfaces (e-mail: jiayizhang@bjtu.edu.cn).

\textbf{Hongyang Du} is currently pursuing the Ph.D. degree with the School of Computer Science and Engineering, Energy Research Institute @ NTU, Nanyang Technological University, Singapore, under the Interdisciplinary Graduate Program. His research interests include semantic communications, resource allocation, and communication theory. (e-mail: hongyang001@e.ntu.edu.sg).

\textbf{Dusit Niyato} [F'17] is a Professor with the School of Computer Science and Engineering, Nanyang Technological University, Singapore. His research interests are in the areas of the Internet of Things (IoT), machine learning, and incentive mechanism design (e-mail: dniyato@ntu.edu.sg).

\textbf{Bo Ai} [F'22] is a Professor with the State Key Laboratory of Rail Traffic Control and Safety, Beijing Jiaotong University, Beijing, China. His research interests include rail traffic mobile communications and channel modeling (e-mail: boai@bjtu.edu.cn).

\textbf{M{\'e}rouane Debbah} [F'15] is a Professor at Khalifa University of Science and Technology in Abu Dhabi. His research interests lie in fundamental mathematics, algorithms, statistics, information, and communication sciences research. He is an IEEE Fellow, a WWRF Fellow, a Eurasip Fellow, an AAIA Fellow, an Institut Louis Bachelier Fellow and a Membre {\'e}m{\'e}rite SEE (e-mail: merouane.debbah@ku.ac.ae).

\textbf{Khaled B. Letaief} [F'03] has been with HKUST since 1993 where he was the Dean of Engineering, and is now a Chair Professor and the New Bright Professor of Engineering. From 2015 to 2018, he was with HBKU in Qatar as Provost. He is an ISI Highly Cited Researcher. He has served in many IEEE leadership positions including ComSoc President, Vice-President for Technical Activities, and Vice-President for Conferences (e-mail: eekhaled@ust.hk).

\end{document}